\documentclass{IEEEtran}
\usepackage{cite}
\usepackage{amsmath,amssymb,amsfonts}
\usepackage{algorithmic}
\usepackage{graphicx}
\usepackage{textcomp}
\usepackage{subfig}
\usepackage{xcolor}
\usepackage{soul}

\def\BibTeX{{\rm B\kern-.05em{\sc i\kern-.025em b}\kern-.08em
    T\kern-.1667em\lower.7ex\hbox{E}\kern-.125emX}}
\begin{document}
\title{Wide-Band/Angle Blazed Dual Mode Metallic Groove Gratings}
\author{Omid Hemmatyar\texttt{$^{\dag}$}, Mohammad Ali Abbassi\texttt{$^{\dag}$}, Babak Rahmani, Mohammad Memarian, and Khashayar Mehrany
\thanks{M. A. Abbassi, M. Memarian and K. Mehrany are with the Electrical Engineering Department, Sharif University of Technology, Tehran 11155-4363, Iran (email:mohammadali.abbassi@ee.sharif.edu; mmemarian@sharif.edu; mehrany@sharif.edu).}

\thanks{B. Rahmani is with the School of Engineering, École Polytechnique
Fédérale de Lausanne (EPFL), 1015 Lausanne, Switzerland (email:babak.rahmani@epfl.ch).}

\thanks{O. Hemmatyar is with the School of Electrical and Computer Engineering, Georgia Institute of Technology, 778 Atlantic Drive NW, Atlanta, GA 30332, USA (email:omid.hemmatyar@gatech.edu).}

\thanks{\dag~These authors contributed equally to this work.}}

\maketitle

\begin{abstract}
We introduce a new and simple approach to acquire wide-band/angle blazing operation in one-dimensional metallic gratings supported by equivalent circuit analysis. The gratings that are investigated here are single-groove gratings that support two propagating guided modes. It is shown that under such operating conditions, one can achieve blazing over a wide range of frequencies and angles. Parameters of the proposed equivalent circuit model for the case of the transverse magnetic (TM) polarization are analytically derived. The accuracy of the models is verified through comparison against results of full-wave simulations. The procedure to achieve wide-band/angle blazing performance is delineated and the design parameters are explicitly given. It is shown that this structure is able to strongly transfer the power of the transverse magnetic (TM) polarized incident wave to the $-1$-th diffraction order with a fractional bandwidth of $\approx50\%$ at working frequency $f_0=10$ GHz and for $–10$dB specular reflection loss, thus realizing a broadband structure. From the circuit model, an insightful discussion is presented at various mechanisms at play, when Bragg and off-Bragg blazing occurs and fully justifies the behavior of the device. The proposed technique opens up new vistas in a wide range of applications such as spectroscopy, Littrow cavities, beam splitters, and frequency scanned antenna reflectors.
\end{abstract}

\begin{IEEEkeywords}
metallic grating, circuit model, blazing operation, Bragg line\end{IEEEkeywords}

\section{Introduction}
Controlling the propagation and diffraction of electromagnetics wave using non-planar and planar structures have been the subject of many studies up to date \cite{johansson1989frequency}. Blazing, the scattering of the diffracted light back in the direction of the incident wave, is an example of light propagation control, used in a plethora of applications such as frequency scanned antenna reflectors, external cavity lasers, spectroscopy, Littrow mounts and multiplexers/demultiplexers \cite{johansson1989frequency}-\nocite{chen2013off}\cite{boyd1995high}.\par
With the advent of artificially engineered three-dimentional structures known as metamaterials and their two-dimensional counterparts (metasurfaces), unprecedent control over the propagation of light has been achieved \cite{alu2005achieving}-\nocite{abdollahramezani2015analog,abdollahramezani2015beam,arik2017polarization,abdollahramezani2017dielectric,abdollahramezani2018reconfigurable,kiarashinejad2019deep,kiarashinejad2019deep_re,hemmatyar2019full,kiarashinejad2019knowledge}\cite{sun2012high}. However, metal losses in metamatrials/metasurfaces and challenges arising in their fabrication process due to their subwavelength lattice constant considerably decrease the efficiency of these structures \cite{monticone2014metamaterials}, \cite{kildishev2013planar}. Thus, alternative structures with high efficiency and ease in manufacturing are highly desirable.\par
Echelette \cite{itoh1969analytical}, \cite{kleemann2012perfect} sinusoidal \cite{maystre1981gratings} and rectangular \cite{hessel1975bragg}, \cite{jull1977gratings} gratings are among various non-planar geometries which are well-known to support blazing operation. Several configurations of reflector-backed dielectric loaded period strip gratings have also been brought forth to obtain perfect blazing effect \cite{jose1987reflector}, \cite{kalhor1988electromagnetic}. Geometrical parameters of these structures are designed to maximize the working bandwidth. The design procedure in these structures is carried out using either a phenomenological \cite{maystre2013diffraction} or numerical approach. Mode matching  and complex value asymptotic methods are among widely employed techniques which obviously require a considerable processing resources. Circuit models are typically a more efficient and faster solution which have been extensively used for the analysis of periodic structures, such as one- and two-dimensional arrangement of cut-through grooves/strips \cite{yarmoghaddam2014circuit}- \nocite{medina2010extraordinary, rodriguez2012analytical, rodriguez2015analytical}\cite{molero2016dynamical}. In these models, however, one does not have direct access to non-specular scattering coefficients and is instead required to calculate specular coefficients as an additional step before retrieving the sought-after non-specular scattered waves.\par
In this paper, the problem of the scattering from one-dimensional metallic gratings when the structure is illuminated by a transverse magnetic (TM) polarized plane wave is revisited in the context of blazing operation. We  investigate the blazing operation in the conventional periodic arrangement of one-dimensional grooves carved in metals using an equivalent circuit model. It is shown that wide-band/angle blazing can be acheieved when the groove width is sufficiently large to support two propagating guided mode inside the grooves.\par
 An acoustic \cite{dong2017broadband} as well as a back-covered grating structure \cite{orazbayev2017wideband} has employed similar approach to increase the bandwidth of the negative refraction and backscattering reduction, respectively. Nontheless, an ultra wide-band structure with low insertion loss, on the one hand, and a simple structure with the smallest possible number of grooves (two) on the other hand has yet to be proposed. Our goal here is to provide a systematic yet simple approach/configuration for achieving wide-band and wide-angle blazed structures. Very recently, a circuit model accounting for the working principles of planar resonant blazed gratings has been presented \cite{memarian2017wide}. To the best of authors' knowledge, no similar model for their non-planar counterparts has been introduced heretofore. Thus, by resorting to the mode matching technique, we analytically solve the electromagnetics waves in different regions and obtain the elements of a circuit model which can \emph{directly} mimic both zero-th and $-1$-th diffraction orders and thus, is usable to study blazing operation. The proposed model for the single-groove gratings consists of two coupled transmission lines for each propagating mode inside the grooves. It should be noted that this is the first time a circuit model capable of modeling two waveguide modes in the groove gratings is proposed. This provides a clear advantage over similar existing models \cite{bagheri2017effect,hemmatyar2017phase,rahmani2018modeling} which employ single mode approximation inside the grooves, enabling us to reach a more accurate solution to the problem. The validity of our model is also verified by full-wave simulations.\par
The paper is organized as follows: Section II is devoted to detailed description of the proposed circuit models. In section III, numerical examples are given to demonstrate the validity of the models. Conclusions are drawn in section III. A time dependence of the form $e^{j\omega t}$ is assumed throughout this paper.
\section{Circuit model}
\begin{figure}[t]
\centering
\includegraphics[width=8.6cm]{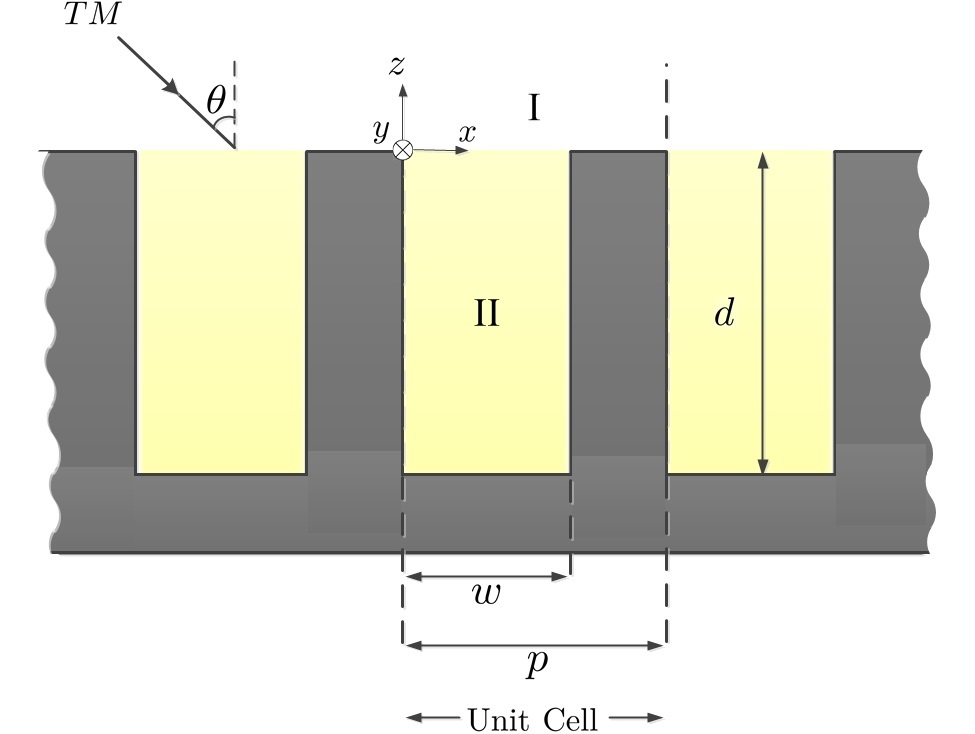}
\caption{\label{Fig.1} One-dimensional semi-infinite metallic grating with one groove per period illuminated by obliquely incident TM polarized plane waves with incident angle $\theta$ with respect to the $z$-direction}
\end{figure}

 Figure \ref{Fig.1} shows the schematic of the structure which is periodic in the $x$-direction with period $ p$, and is extended to infinity in the $y$-direction. The grooves have widths $w$ and hieght $d$ which are filled with a dielectric medium with refractive index $n_2$ (region II). The refractive index of the homogeneous medium above the grating (region I) is $n_1$.\par
\begin{figure}[t!]
\centering
\includegraphics[trim={0cm 1cm 0cm 0cm},,width=0.48\textwidth]{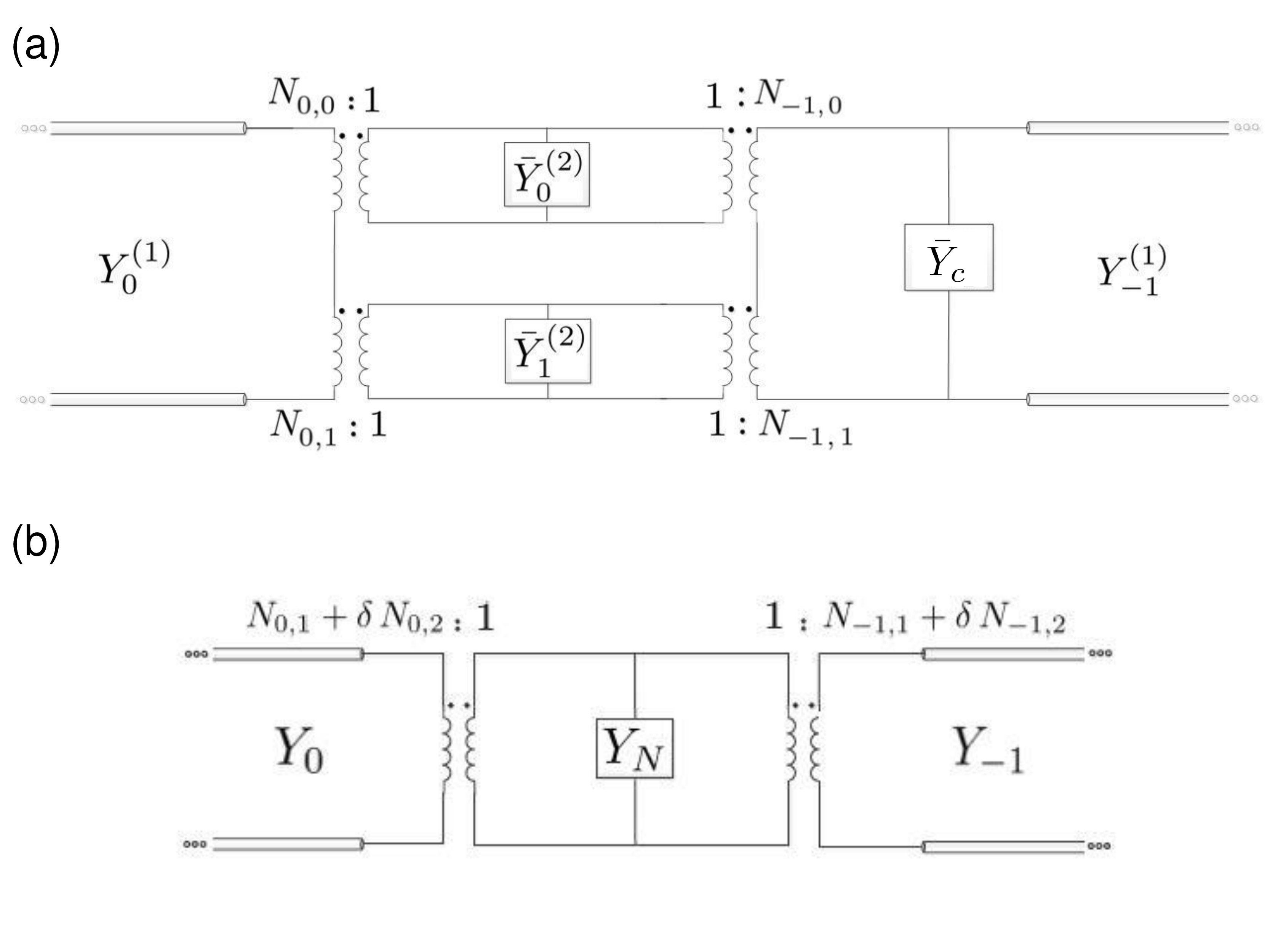}
\caption{(a) The equivalent circuit model derived for the unit cell of the reflection structure
shown in Fig. \ref{Fig.1}.(b) Simplified circuit model under the condition that the incident wave only excites the $0$-th order diffracted port.}
\label{Fig.2}
\end{figure}
The structure is illuminated by an obliquely incident TM polarized plane wave (the magnetic field in the $y$-direction) with incident angle $\theta$ with respect to the $z$-direction. The tangential magnetic and electric fields in region I can be written as:
\begin{subequations}
\begin{equation}\label{1a}
H_{1y} = H_{1,0}^{+} e^{j k_{z_0} z} e^{-jk_{x_0} x} +  \sum \limits_{n} {H_{1,n}^{-} e^{-j k_{z_n} z} e^{-jk_{x_n} x}},
\end{equation}
\begin{equation}\label{1b}
E_{1x} = -\xi_{1,0} H_{1,0}^{+} e^{j k_{z_0} z} e^{-jk_{x_0} x}  +  \sum \limits_{n} {\xi_{1,n} H_{1,n}^{-} e^{-j k_{z_n} z} e^{-jk_{x_n} x}}.
\end{equation}
\end{subequations}
Here, $k_{x_0}=k_0 n_1 \sin\theta$ is the transverse wavevector of the incident electromagnetic field in which $k_0$ is the free space wavenumber, and $k_{x_n}$ and $k_{z_n}$ are the wavevector components of the $n$-th order diffracted electromagnetic fields  in region I, given by:
\begin{subequations}
\begin{equation}\label{2a}k_{x_n} = k_{x_0} + \frac{2n \pi}{p},\end{equation}
\begin{equation}\label{2b}k_{z_n} = \sqrt{k_0^2 n_1^2 - k_{x_n}^2}.\end{equation}
\end{subequations}
It should be noted that the positive real value and the negative imaginary value of $k_{zn}$ correspond to the propagating and evanescent waves, respectively. Moreover, $\xi_{1n} = k_{zn}/ \left( \omega \varepsilon_0 n_1^2 \right)$ is the TM-wave admittance of the $n$th diffracted order in region I.\par
In region II, we assume that the grooves support the first two waveguide modes, i.e., we take into account the fundamental transverse electromagnetic (TEM) and transverse magnetic (TM)$_1$ modes which are propagating inside the grooves. The validity of this approximation is limited to the frequency range of $ w n_2<\lambda_0$ where the higher modes inside the groove are evanescent. As a result, the magnetic and electric fields inside the groove can be written in the form:
\begin{subequations}
\begin{align}\label{3a} H_{2y} =& H_{2,0}^+ e^{j \beta_{0} z} + H_{2,0}^- e^{-j \beta_{0} z}
  +\nonumber\\& H_{2,1}^+ \cos \left( \frac{\pi x}{w} \right) e^{j \beta_{1} z}+ H_{2,1}^- \cos \left( \frac{\pi x}{w} \right) e^{-j \beta_{1} z},
\end{align}
\begin{align}\label{3b}
E_{2x} = &- \xi_{2,0} H_{2,0}^+ e^{j \beta_{0} z} + \xi_{2,0} H_{2,0}^- e^{-j \beta_{0} z}
-\nonumber\\ & \xi_{2,1} H_{2,1}^+ \cos \left( \frac{\pi x}{w} \right) e^{j \beta_{1} z} + \xi_{2,1} H_{2,1}^- \cos \left( \frac{\pi x}{w} \right) e^{-j \beta_{1} z}
\end{align}
\end{subequations}
for $x \in \left[0,w \right]$, where 
\begin{equation}\label{4}
\beta_{0} = k_0 n_2, \qquad  \beta_{1} = \sqrt{ (k_0 n_2)^2 - \left( \frac{\pi}{w} \right)^2 }
\end{equation}
are the propagation constants of the TEM and TM$_1$ modes supported by a parallel plate waveguide, and 
\begin{equation}\label{5}
\xi_{2,i} = \frac{\beta_{i}}{\omega \varepsilon_0 n_2^2}, \quad i=0,1
\end{equation}
is the wave admittance of the $i$-th mode inside each groove.\par

Following the procedure explained in the Appendix, one can easily obtain an equivalent circuit model for this structure. Figure \ref{Fig.2}.(a) shows the schematic of this circuit model which is composed of three different admittances as follows:
\begin{subequations}
\begin{equation}\label{6a}
\bar{Y}_{0}^{(2)} ={Y}_{0_L}^{(2)}-|N_{-1,0}|^2 \bar{Y}_c+\sum \limits_{n \ne -1,0} |N_{n,0}|^2{Y_n^{(1)} },
\end{equation}
\begin{equation}\label{6b}
\bar{Y}_{1}^{(2)} ={Y}_{1_L}^{(2)}-  |N_{-1,1}|^2 \bar{Y}_c+  \sum \limits_{n \ne -1,0} |N_{n,1}|^2 {Y_n^{(1)} } ,
\end{equation}
\begin{equation}\label{6c}
\bar{Y}_c = \sum \limits_{n \ne -1,0} {\frac{N_{n,1} N_{n,0}^\ast}{N_{-1,1} N_{-1,0}^\ast} Y_n^{(1)} }.
\end{equation}
\end{subequations}
Here, 
$Y_n^{(1)}=\frac{1}{p \, \xi_{1,n}}$ is the characteristic admittance of the $n$-th diffracted order, and
\begin{subequations}
\begin{equation}\label{7a}Y_{0_L}^{(2)}=\frac{-j\cot{\beta_0 d}}{w \xi_{2,0}},\end{equation}
\begin{equation}\label{7b}Y_{1_L}^{(2)}=\frac{-j\cot{\beta_1 d}}{2 w \xi_{2,1}}\end{equation}
\end{subequations}
are the input admittance of the TEM and TM$_1$ modes inside the grooves, respectively. Moreover, $N_{n,0}$ and $N_{n,1}$ are the coupling between the $n$-th diffracted order and each modes inside the groove, which are given by
\begin{subequations}
\begin{equation}\label{8a}N_{n,0}=e^{j k_{x_n} \frac{w}{2}} \, \textrm{sinc} \left( \frac{k_{x_n} w}{2} \right),\end{equation}
\begin{equation}\label{8b}N_{n,1}=\frac{2j k_{x_n}}{w\left( k_{x_n}^2 - \left( \frac{\pi}{w} \right)^2\right)}\cos \left( \frac{k_{x_n} w}{2} \right)e^{ j k_{x_n} \frac{w}{2}}.\end{equation}
\end{subequations}
\par
Since the incident wave solely excites the $0$-th order diffracted port, the circuit model can be simplified to Fig. \ref{Fig.2}.(b), in which
\begin{equation}\label{9}\delta=\frac{N_{0,1}^\ast \bar{Y}_0^{(2)}+N_{-1,0}\left(N_{0,1}^\ast N_{-1,0}^\ast-N_{0,0}^\ast N_{-1,1}^\ast\right)\left(Y_{-1}^{(1)}+\bar{Y}_c\right)}{N_{0,0}^\ast \bar{Y}_1^{(2)}+N_{-1,1}\left(N_{0,0}^\ast N_{-1,1}^\ast-N_{0,1}^\ast N_{-1,0}^\ast\right)\left(Y_{-1}^{(1)}+\bar{Y}_c\right)}\end{equation}
and
\begin{equation}\label{10}Y_N=\frac{P_0^\ast}{N_{0,0}^\ast}\left[\bar{Y}_{0}^{(2)}+N_{-1,0}^\ast P_{-1}\left(Y_{-1}^{(1)}+\bar{Y}_c\right) \right]-|P_{-1}|^2 Y_{-1}^{(1)}\end{equation}
where $P_i=N_{i,0}+\delta N_{i,1}$. Now, the $0$-th and $-1$-th order reflection coefficients can be derived from the circuit model, which are given by
\begin{subequations}
\begin{equation}\label{11a}\Gamma_0=\frac{Y_0^{(1)}-\frac{1}{|P_0|^2}\left(Y_N+|P_{-1}|^2 Y_{-1}^{(1)}\right)}{Y_0^{(1)}+\frac{1}{|P_0|^2}\left(Y_N+|P_{-1}|^2 Y_{-1}^{(1)}\right)},\end{equation}
\begin{equation}\label{11b}\Gamma_{-1}=\frac{P_{-1}}{P_0}(1+\Gamma_0).\end{equation}
\end{subequations}

We will use the results obtained in this section for studying the blazing operation in the next section.

\section{Blazing condition}
The diffraction of an obliquely incident wave on a periodic grating follows the Bragg relation which reads as $\sin(\theta_i)+\sin(\theta_m)=m\lambda_0/p$. Here the period of the grating $p$, the incident angle $\theta_i$, the order of the diffraction $m$ and its corresponding reflection angle $\theta_m$ and the working wavelength in the free space $\lambda_0$ are related. For autocollimation ($\theta_i=\theta_m$) to occur with the $m=-1$-th diffracted wave (i.e. $\theta_i = \theta_{-1}$ as needed in typical Littrow mounts), the period is dictated by the Bragg relation as
\begin{equation}\label{12}
  p=\lambda_0/2\sin(\theta_0).
\end{equation}
Although (\ref{12}) guarantees the propagation of the $-1$-th diffraction order in the path of the incident wave, yet it does not comment on the amplitudes of $m=0$-, $ m=-1$-th and higher diffraction orders. Nevertheless, the region of interest in our work is where only $m=0$- and $ m=-1$-th orders are propagating (i.e., positive real value of $\xi_{1n}$ for $n = 0,1$) while all other higher diffraction orders are evanescent (i.e., imaginary negative value of $\xi_{1n}$ for $n \ne 0,1$). In such a case, $Y_0^{(1)}$ and $Y_{-1}^{(1)}$ will be real while the characteristics admittances of all other diffracted orders will be pure imaginary which make the $Y_N$ admittance in the circuit model shown in Fig. \ref{Fig.2}.(b) to be reactive. If we neglect the impact of all the evanescent diffracted orders ($\bar{Y}_c = 0$), the blazing occurs at the Fabry-Perot resonance of the groove's modes (i.e., when $d \simeq m \, \lambda/2$ ($d \simeq (m+0.5) \, \lambda/2$) for constructive (destructive) interference between incident and reflected waves, where $m$ is an integer). Thus, according to the circuit model shown in Fig. \ref{Fig.2}.(b), the blazing can be achieved when $Y_N=0$, and
\begin{equation}\label{13}|N_{0,1}+\delta N_{0,2}|^2 Y_0^{(1)}=|N_{-1,1}+\delta N_{-1,2}|^2 Y_{-1}^{(1)}\end{equation}
which are the generalization of the resonance, and Bragg conditions, respectively. When these conditions are established, the specular reflection will vanish, and the incident power will be wholly reflected in the $-1$-th diffracted order.\par
In such a structure, the only adjustable parameters, in addition to the materials inside the grooves which we assume to be fixed, are the height and width of the grooves. The former determines the first Fabry-Perot resonance of the groove's modes, which will be located at $f\simeq10.7$ GHz for $d=7$ mm when the destructive interference between incident and reflected waves occur (i.e. $d \simeq (m+0.5) \, \lambda/2$ or $f=(m+0.5)c/2d$ for $m=0$). The latter determines the cut-off frequency of the second mode inside the grooves and its propagating constant. Then, four different scenarios are conceivable based on the frequency of the Fabry-Perot resonance of the TEM groove's mode, as follows:
\subsection{Extremely under the cut-off}
\begin{figure}[t!]
\centering
\includegraphics[trim={0cm 0cm 0cm 0cm},width=0.5\textwidth, clip]{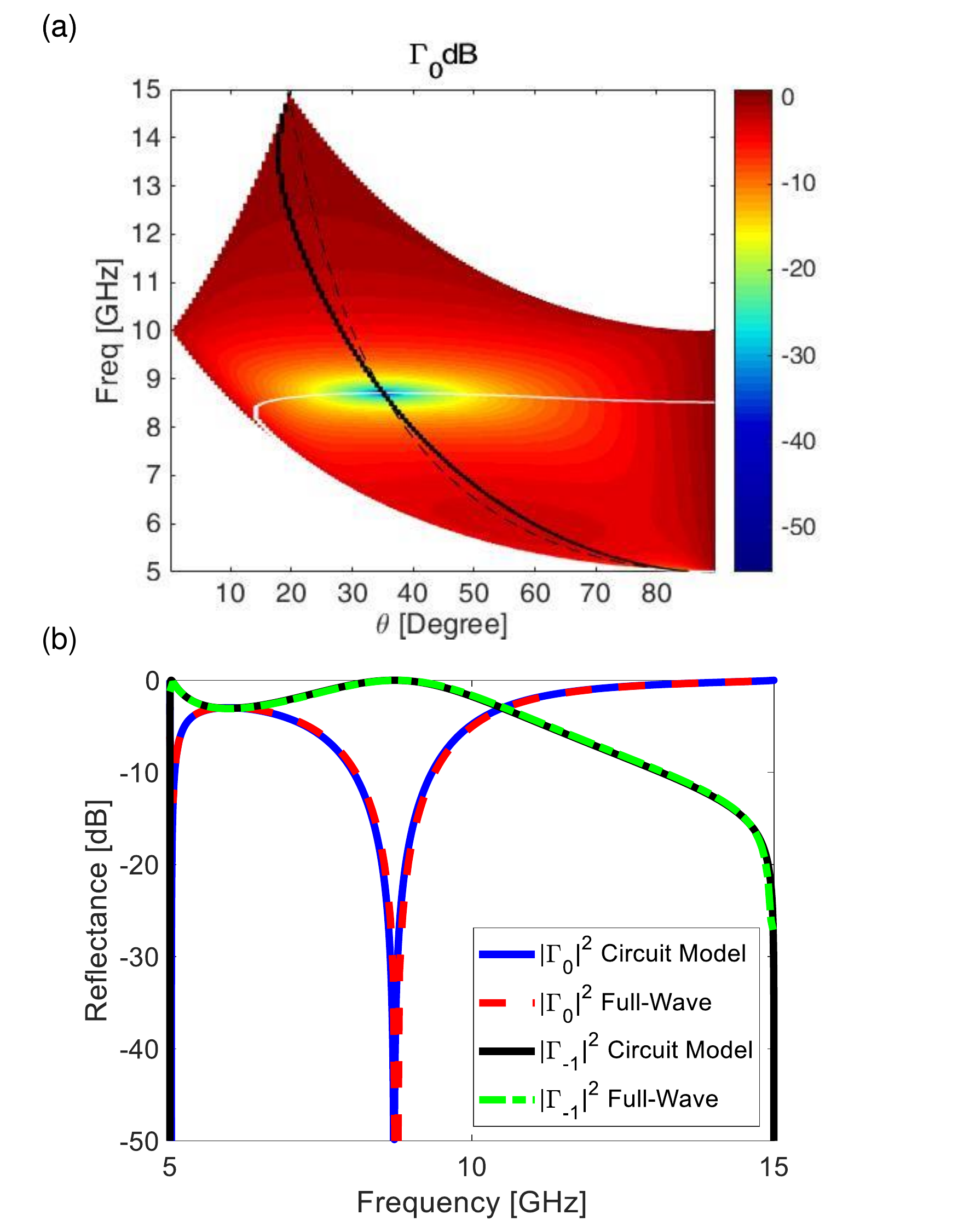}
\caption{\label{Fig.ft1} (a)The specular reflected power versus the frequency and incident angle for $w=5$ mm and  groove using the circuit model. Other parameters are  $n_1= n_2=1$, $p = 30$ mm, $d = 7$ mm. The dashed black line is the general Bragg condition, the solid black line is the modified-Bragg condition (i.e. (\ref{13})), and the solid white line is the resonance condition (i.e. $Y_N=0$). (b) The comparison between reflectance spectrum (when moving on the Bragg line, i.e., dashed black line in (a)) obtained from our equivalent circuit model with those from full-wave simulation.}
\end{figure}
\begin{figure}[t!]
\centering
\includegraphics[trim={0cm 0cm 0cm 0cm},width=0.5\textwidth, clip]{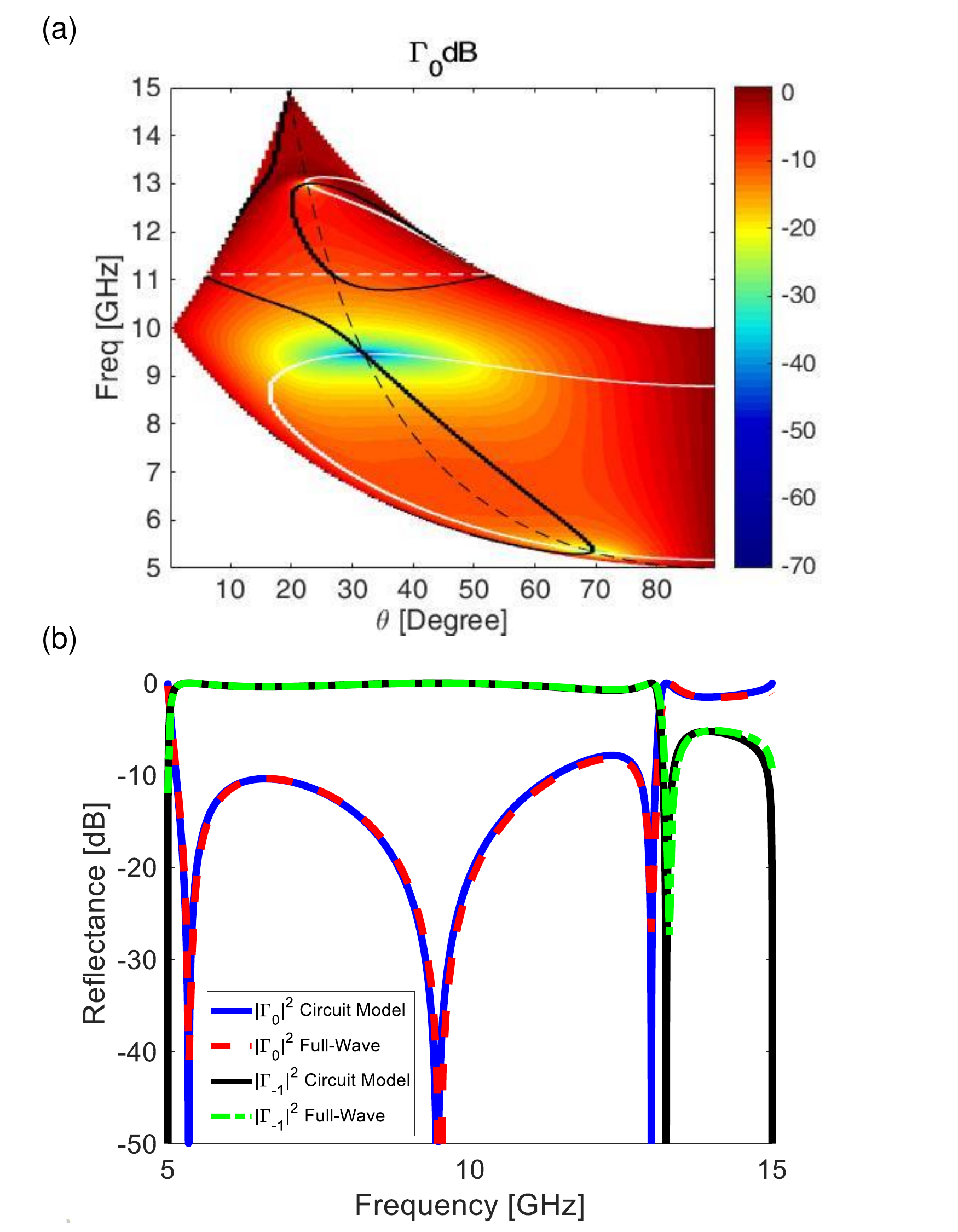}
\caption{\label{Fig.ft2} (a) The specular reflected power versus the frequency and incident angle for $w=13.5$ mm and  groove using the circuit model. Other parameters are  $n_1= n_2=1$, $p = 30$ mm, $d = 7$ mm. The dashed black and white lines show the general Bragg condition and the cut-off for the second mode (TM$_1$) inside the groove, respectively. The solid black line is the modified-Bragg condition (i.e. (\ref{13})), and the solid white line is the resonance condition (i.e. $Y_N=0$). (b) The comparison between reflectance spectrum (when moving on the Bragg line, i.e., dashed black line in (a)) obtained from our equivalent circuit model with those from full-wave simulation.}
\end{figure}

When $w$ is sufficiently small, the Fabry-Perot resonance of the groove's TEM mode lies extremely under the cut-off frequency of the TM$_1$ one. In such a case, $\delta\approx 0$ and the impact of the second mode inside the groove on the blazing can be almost neglected. Therefore, it is expected to observe the blazing on the Bragg line and when the short-circuited TEM mode inside the groove resonates with all the evanescent diffracted orders, i.e., $\bar{Y}_{0_L}^{(2)}=\sum \limits_{n \ne -1,0} |N_{n,0}|^2{Y_n^{(1)} }$. Furthermore, blazing can also be observed when $\theta \to \pi/2$ on the Bragg line since $Y_0^{(1)}$ and $Y_{-1}^{(1)}$ approach infinity (see Fig.~\ref{Fig.ft1}.(a)).\par
Fig.~\ref{Fig.ft1}.(a) shows the specular reflected power versus the frequency and incident angle for $w=5$ mm, $p=30$ mm and $d=7$ mm in the regime where the $0$-th and $-1$-st diffracted orders are solely propagating. The dashed black line is the general Bragg condition (i.e. $\sin(\theta_i)+\sin(\theta_m)=m\lambda_0/p$). The solid black line represents the modified-Bragg condition (i.e. when the Bragg condition for a single-mode groove is modified by (\ref{13})), and the solid white line shows the resonance condition ($Y_N=0$). These last two conditions together satisfy the matching condition, so we expect that wherever the solid black line and the solid white line cross each other, all the power of the TM-polarized incident wave transfers to the $-1$-st diffraction order with a fractional bandwidth at the working frequency of $f_0 \simeq 8.75$ GHz, and the specular reflection vanishes. As expected, the specular reflection vanishes near the resonance of the TEM mode on the Bragg line (see Fig.~\ref{Fig.ft1}.(a)). It should be noted that the difference between this resonance with the Fabry-Perot resonance (i.e. $f=c/4d \simeq 10.7$ GHz) can be ascribed to the modification imposed by (\ref{13}). In such a case, where the second mode inside the groove is extremely under the cut-off, the fractional bandwidth of the blazing is about $14\%$. To evaluate the accuracy of our presented circuit model, the reflectance spectrum versus the frequency is plotted in Fig.~{\ref{Fig.ft1}}.(b) by moving on the Bragg line (dashed black line in Fig.~{\ref{Fig.ft1}}.(a)). A good agreement observed between the equivalent circuit model results with those obtained from full-wave simulations.

\begin{figure}[t!]
\centering
\includegraphics[trim={0cm 0cm 0cm 0cm},width=0.5\textwidth, clip]{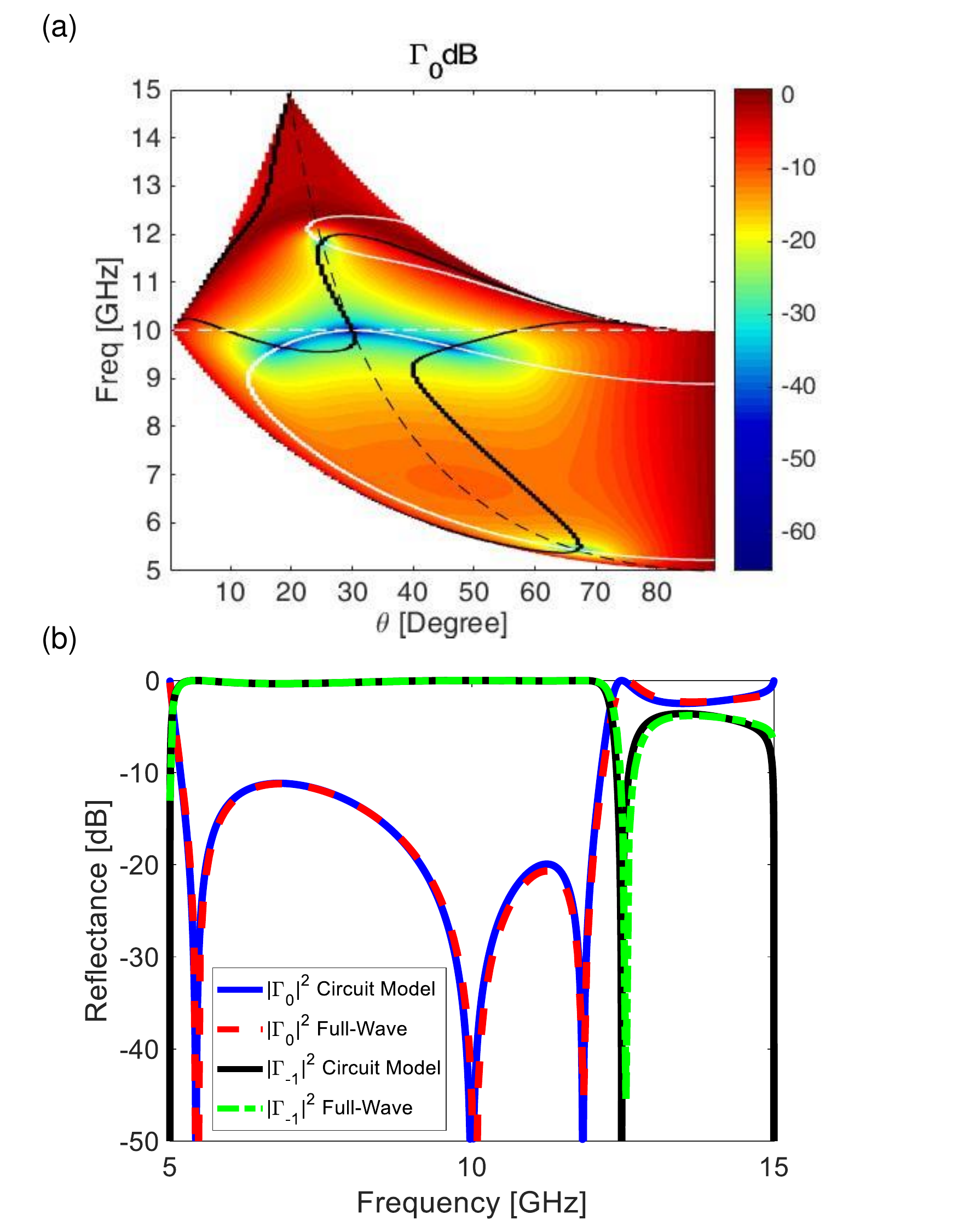}
\caption{\label{fig.ft3} (a)The specular reflected power versus the frequency and incident angle for $w=15$ mm and  groove using the circuit model. Other parameters are  $n_1= n_2=1$, $p = 30$ mm, $d = 7$ mm. The dashed black and white lines show the general Bragg condition and the cut-off for the second mode (TM$_1$) inside the groove, respectively. The solid black line is the modified-Bragg condition (i.e. (\ref{13})), and the solid white line is the resonance condition (i.e. $Y_N=0$). (b) The comparison between reflectance spectrum (when moving on the Bragg line, i.e., dashed black line in (a)) obtained from our equivalent circuit model with those from full-wave simulation.}
\end{figure}

\begin{figure}[t!]
\centering
\includegraphics[trim={0cm 0cm 0cm 0cm},width=0.5\textwidth, clip]{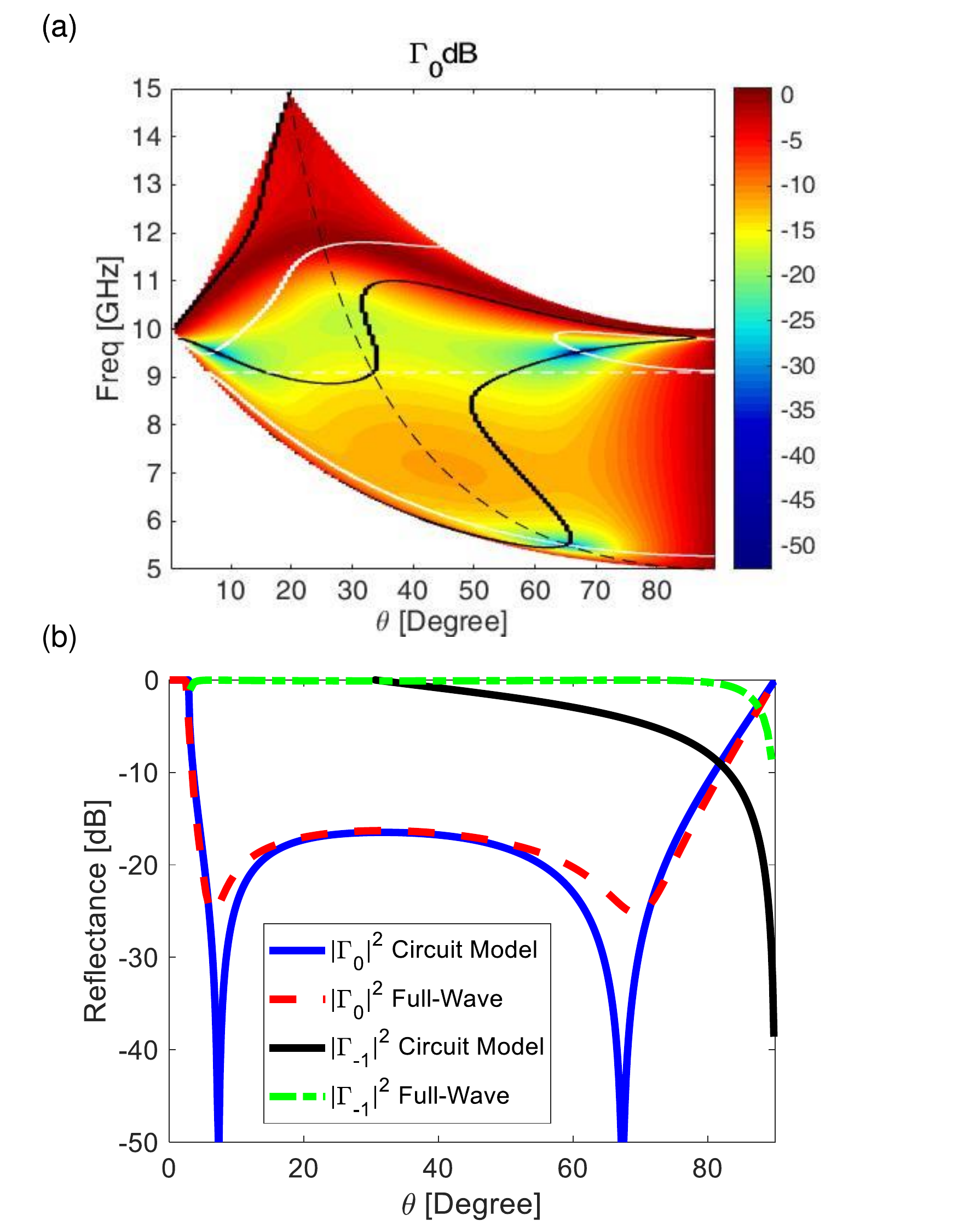}
\caption{\label{fig.ft4} (a) The specular reflected power versus the frequency and incident angle for $w=16.5$ mm and  groove using the circuit model. Other parameters are  $n_1= n_2=1$, $p = 30$ mm, $d = 7$ mm. The dashed black and white lines show the general Bragg condition and the cut-off for the second mode (TM$_1$) inside the groove, respectively. The solid black line is the modified-Bragg condition (i.e. (\ref{13})), and the solid white line is the resonance condition (i.e. $Y_N=0$). (b) The reflected power versus the incident angle for $w=16.5$ mm at $f = 9.52$ GHz. Other parameters are same to those in (a).}
\end{figure}

\subsection{Under the cut-off}
If we increase the width of the grooves, the role of the second mode inside the grooves no longer can be neglected. First, assume that the Fabry-Perot resonance of the groove's TEM mode lies below the cut-off frequency of the second mode. Fig.~\ref{Fig.ft2}.(a) shows the specular reflected power for such a scenario, where $w=13.5$ mm, $p=30$ mm, $d=7$ mm, and the dashed white line shows the cut-off for the second mode (TM$_1$) inside the groove. According to this figure, the specular reflection has three zeros on the Bragg line (i.e. the intersections of solid black and white lines, where the perfect matching condition is satisfied). Two of them are due to the resonance of the TEM and TM$_1$ modes inside the grooves. The other is the horizontal blazing seen at $\theta \to \pi/2$ which is slightly shifted due to the presence of the second mode inside the grooves. Fig.~{\ref{Fig.ft2}}.(b) shows that our presented circuit model can predict these three peaks in the reflectance spectrum when moving on the Bragg line (dashed black line in Fig.~{\ref{Fig.ft2}}.(a)).
\subsection{Near the cut-off}
By increasing the width of the grooves such that the Fabry-Perot resonance of the TEM mode lies near the cut-off frequency of the TM$_1$ mode, off-Bragg blazings emerge in addition to the conventional Blazing on the Bragg line. This is clearly shown in Fig.~\ref{fig.ft3}.(a), where the specular reflection power is depicted for for $w=15$ mm, $p=30$ mm, $d=7$ mm. In such a case, both the Fabry-Perot resonance of the TEM mode and the cut-off frequency of the TM$_1$ mode are located near $f=10.7$ GHz. Fig.{~\ref{fig.ft3}}.(b) shows a good match between our equivalent circuit model and full-wave simulation results. Comparing the reflectance spectrum shown in Fig.{~\ref{Fig.ft2}.(b) with that in Fig.{~\ref{fig.ft3}}.(b), one can see that the blazing bandwidth (the spectral range where $\Gamma_{-1} \simeq 0$ dB, and $\Gamma_0 < -10$ dB) is increasing as we are getting close to the cut-off frequency of TM$_1$ mode inside the groove.}

\subsection{Above the cut-off}
Fig.~\ref{fig.ft4}.(a) shows the specular reflection when the Fabry-Perot resonance of the TEM mode lies above the cut-off frequency of the TM$_1$ mode, where $w=16.5$ mm, $p=30$ mm, $d=7$ mm. In such a case, the blazing condition is not satisfied on the Bragg line except the horizental blazing, and the groove's resonances solely provide the off-Bragg blazing. This results in a blazing in the reflectance response with a wide bandwidth in both frequency and angle (see Fig.~{\ref{fig.ft4}.(b)}).

\section{Conclusion}
In this work, non-planar one-dimensional reflected-back simple (one groove per period) grating was studied to achieve wide-band blazing operation. By employing the mode matching technique, we developed an equivalent circuit model to mimic simple grating by considering two guided modes inside the grooves. Accuracy of the proposed circuit models is verified by comparing circuit results with those of the full-wave simulations. The width and height of the grooves in the simple grating structure are designed insomuch two waveguide modes propagate and couple to each other within the grooves. The ultra blazing bandwidth in this case is due to coupling of the two guided modes inside the grooves. By taking advantage of the presented approach, one can easily consider even more propagating modes inside the groove to improve the accuracy of their design. Moreover, this approach can be extended to more complex structures comprising of more than one groove inside each period, where the coupling of evanescent modes inside the adjacent grooves at the surface has not been investigated so far. Therefore, the presented approach can be employed to design more complex gratings at appropriate frequency for different applications such as antenna reflectors and lasers. In addition, to accelerate the design and optimization processes of such gratings with more complicated physics on one hand, and extract valuable understanding about the effect of the coupling between the evanescent modes inside the adjacent grooves on the blazing condition on the other hand, recent deep-learning based approaches, such as dimensionality reduction \cite{kiarashinejad2019deep_re,kiarashinejad2019deep} and knowledge discovery \cite{kiarashinejad2019knowledge} can be utilized.

\appendix*
\section{}
Applying the boundary conditions at $z=0$ for the electric fields, i.e., the continuity of the tangential electric field at every point of the unit cell, leads to the following equations:
\begin{subequations}
\begin{align}\label{14a}
\left( H_{1,0}^+ - H_{1,0}^- \right) p \xi_{1,0} &= \left( H_{2,0}^+ -H_{2,0}^-\right)  w \xi_{2,0} N_{0,0} 
\nonumber\\&+ \left(H_{2,1}^+ -H_{2,1}^- \right)w \xi_{2,1} N_{0,1},
\end{align}
\begin{align}\label{14b}
 - H_{1,-1}^-  p  \xi_{1,-1} &= \left( H_{2,0}^+ -H_{2,0}^-\right)  w \xi_{2,0} N_{-1,0} 
\nonumber\\&+ \left(H_{2,1}^+ -H_{2,1}^- \right) w \xi_{2,1}  N_{-1,1},
\end{align}
\begin{align}\label{14c}
 - H_{1,n}^-  p  \xi_{1,n} &=  \left( H_{2,0}^+ -H_{2,0}^-\right)  w \xi_{2,0} N_{n,0} 
\nonumber\\&+ \left(H_{2,1}^+ -H_{2,1}^- \right)  w \xi_{2,1}  N_{n,1} \quad n \ne -1,0
\end{align}
\end{subequations}
where
\begin{subequations}
\begin{align}\label{15a}
N_{n,0} = \frac{1}{w} \int_{0}^{w} {e^{ j k_{x_n} x} dx} 
= e^{ j k_{x_n} \frac{w}{2}} \, \textrm{sinc} \left( \frac{k_{x_n} w}{2} \right),
\end{align}
\begin{align}\label{15b}
N_{n,1} &= \frac{1}{w} \int_{0}^{w} {e^{ j k_{x_n} x} \, \cos \left( \frac{\pi x}{w} \right) \, dx}
\nonumber\\&= \frac{2j k_{x_n}}{w\left( k_{x_n}^2 - \left( \frac{\pi}{w} \right)^2\right)}\cos \left( \frac{k_{x_n} w}{2} \right)e^{ j k_{x_n} \frac{w}{2}}.
\end{align}
\end{subequations}
Next, we apply the boundary condition for tangential magnetic fields at the opening of each groove at $z=0$:
\begin{subequations}
\begin{align}\label{16a}
\left( H_{2,0}^+ +H_{2,0}^-\right) &= \left( H_{1,0}^+ +  H_{1,0}^- \right) N_{0,0}^\ast \nonumber\\&+  H_{1,-1}^- \, N_{-1,0}^\ast
+ \sum \limits_{n \ne -1,0} {H_{1,n}^- N_{n,0}^\ast},
\end{align}
\begin{align}\label{16b}
\frac{1}{2} \left(H_{2,1}^+ +H_{2,1}^-\right)&= \left( H_{1,0}^+ +  H_{1,0}^- \right) N_{0,1}^\ast \nonumber\\&+   H_{1,-1}^-  N_{-1,1}^\ast
+ \sum \limits_{n \ne -1,0} {H_{1,n}^- N_{n,1}^\ast} 
\end{align}
\end{subequations}
where (\ref{16a}) is obtained using (\ref{1a}) and (\ref{3a}) by integrating both sides over the width of grooves II and (\ref{16b}) is obtained using (\ref{1b}) and (\ref{3b}) by multiplying the electric fields by $\cos (\pi x/w)$ and then integrating both sides over the width of grooves II. Furthermore, as the grooves are short-circuited at the distance $d$ from the surface, the forfward and backward fields inside the groove are related, as follows:
\begin{equation}\label{17}H_{2,i}^-=e^{-2j\beta_i d}H_{2,i}^+.\end{equation}\par
Now, we can model the $0$-th and $-1$-th diffracted order with a transmission line whose voltage and current are given by
\begin{subequations}
\begin{equation}\label{18a}V_{i}^{(1)}=\left( H_{1,i}^+ - H_{1,i}^- \right) p  \xi_{1,i},\end{equation}
\begin{equation}\label{18b}I_{i}^{(1)}=H_{1,i}^+ + H_{1,i}^-\end{equation}
\end{subequations}
in which $i\in\{0,-1\}$. Furthermore,
\begin{equation}\label{19a}V_{i}^{(2)}=\left( H_{2,i}^+ - H_{2,i}^- \right) w  \xi_{2,i},\end{equation}
\begin{equation}\label{19b}I_{i}^{(2)}=H_{2,i}^+ + H_{2,i}^-\end{equation}
are the voltage and current of $i$-th groove mode ($i\in\{0,1\}$) at the surface of the grating (i.e. $z=0$). Therefor, we can rewrite (\ref{14a}) and (\ref{14b}) as:
\begin{subequations}
\begin{equation}\label{20a}
V_{0}^{(1)} = N_{0,0} V_{0}^{(2)} + N_{0,1} V_{1}^{(2)},
\end{equation}
\begin{equation}\label{20b}
V_{-1}^{(1)} = N_{-1,0} V_{0}^{(2)} + N_{-1,1} V_{1}^{(2)}
\end{equation}
\end{subequations}
and simplify (\ref{16a}) and (\ref{16b}) as:
\begin{subequations}
\begin{align}\label{21a}
I_{0}^{(2)} &= N_{0,0}^* \,  I_{0}^{(1)} + N_{-1,0}^* \,  I_{-1}^{(1)} \nonumber\\&- \sum \limits_{n \ne -1,0} {\frac{|N_{n,0}|^2}{p \, \xi_{1,n}}} V_{0}^{(2)} 
- \sum \limits_{n \ne -1,0} {\frac{N_{n,1} \, N_{n,0}^\ast}{p \, \xi_{1,n}}} V_{1}^{(2)},
\end{align}
\begin{align}\label{21b}
I_{1}^{(2)}& = N_{0,1}^* \,  I_{0}^{(1)} + N_{-1,1}^* \,  I_{-1}^{(1)}\nonumber\\& - \sum \limits_{n \ne -1,0} {\frac{N_{n,1}^-\ast\, N_{n,0}}{p \, \xi_{1,n}}} V_{0}^{(2)}
-  \sum \limits_{n \ne -1,0} {\frac{|N_{n,1}|^2}{p \, \xi_{1,n}}} V_{1}^{(2)}.
\end{align}
\end{subequations}
Then, according to the above voltage and current expressions we can propose a circuit model for this structure which is depicted in Fig. \ref{Fig.2}.(a).

\bibliographystyle{IEEEtran}

\bibliography{Reference.bib}

\end{document}